# DNA Tweezers Based on Semantics of DNA Strand Graph

MANDRITA MONDAL
Electronics and Communication Science Unit
Indian Statistical Institute
203, B.T Road, Kolkata-700108, India
E-mail: mandritamondal@gmail.com
Tel: +91 9830354798

KUMAR S. RAY
Electronics and Communication Science Unit
Indian Statistical Institute
203, B.T Road, Kolkata-700108, India
E-mail: ksray@isical.ac.in
Tel: +91 8981074174

**Abstract**

Because of the limitations of classical silicon based computational technology, several alternatives to traditional method in form of unconventional computing have been proposed. In this paper we will focus on DNA computing which is showing the possibility of excellence for its massive parallelism, potential for information storage, speed and energy efficiency. In this paper we will describe how syllogistic reasoning by DNA tweezers can be presented by the semantics of process calculus and DNA strand graph. Syllogism is an essential ingredient for commonsense reasoning of an individual. This paper enlightens the procedure to deduce a precise conclusion from a set of propositions by using formal language theory in form of process calculus and the expressive power of DNA strand graph.

## 1. Introduction

In the present world, life can't be imagined without computer. For past several decades conventional silicon based computing has broadly been applied in almost every domain of modern technology. Because of the limitations of classical silicon based computational technology in terms of design complexity, memory requirement, energy consumption, processing power and heat dissipation; we are approaching towards a paradigm shift from silicon to carbon. Several alternatives to traditional method in form of unconventional computing have been proposed. Now a days the unconventional methods of computing, for example, molecular computing, quantum computing, DNA computing, cellular automata and amorphous computing, are gaining popularity and being cultivated widely in computational research. In this paper we will focus on DNA computing which is showing the possibility of excellence for its massive parallelism, potential for information storage, speed and energy efficiency. DNA computing uses

DNA strands and chemical operations for manipulating the strands to perform computation [Adleman, 1994; Winfree *et al.*, 1998;Benenson *et al.*, 2001; Chang and Gou, 2003; Green *et al.*, 2006; Akerkar and Sajja, 2009], logical reasoning and decision making [Yeung and Tsang, 1997; Ray and Mondal, 2011a; Ray and Mondal, 2011b; Ray and Mondal, 2016].

In this paper we will show how the mechanism of DNA tweezers for solving chaining syllogism can be presented by the semantics of process calculus. Finally the graphical depiction of the program is performed using DNA strand graph [Petersen *et. al.*, 2016].

DNA tweezers [Yurke and Mills, 2003] are DNA-fuelled dynamic devices that works according to the principle of toehold mediated branch migration and DNA strand displacement. We have solved reasoning with dispositions [Zadeh, 1985] using DNA tweezers in our paper [Ray and Mondal, 2012]. In this paper we will use DNA tweezers to perform syllogistic reasoning based on a given set of propositions. Syllogism is a logical inference mechanism which generates conclusion from two or more propositions by deductive reasoning. The syllogistic reasoning generally contains three propositions which are major premise, minor premise and conclusion. When a chain of conditional statements i.e. premises are used to deduct conclusion, it is called chaining syllogism.

In this paper the formal language theory is used as a tool to model and analyze the biochemical reactions which are required to solve chaining syllogism problems by DNA tweezers. To define the semantics of the procedures performed in wet lab and to formalize the architecture of the entire model of reasoning, formal language theory is required. To build the model and to simulate and analyze DNA strand displacement systems, a domain-specific DNA strand displacement (DSD) language was developed [Lakin *et. al.*, 2012; Phillips and Cardelli, 2009].Different complex DNA models are being used for computation [Adleman, 1994], reasoning [Ray and Mondal, 2011a] and classification [Ray and Mondal, 2011b] by DNA computing. But DSD language can only define and analyze the primary structures of DNA strands. Thus, redefinition and extension of present DSD language is needed. Petersen *et. al.* [Petersen *et. al.*, 2016] proposed a reformulated language, termed as *process calculus*. The displacement mechanism of DNA strands with rich secondary structures can successfully be modeled, simulated and analyzed by the expressive syntax and formal semantics of newly proposed formal language, process calculus. A graphical representation is required for better understanding of biochemical reactions of complicated formal DNA models for solving computation, classification and reasoning problems. The graphical depiction of process calculus is termed as *strand graph* [Petersen *et. al.*, 2016].

## 2. DNA-tweezers

*DNA tweezers*, first demonstrated by Yurke *et al.* [Yurke and Mills, 2003], is a molecular device which follows the principle of *toehold mediated DNA strand displacement*. The DNA strands works as structural material as well as the 'fuel' of this dynamic device. A set of tweezers is made up of three single stranded DNA sequences. The DNA tweezers have two partially double stranded arms which are connected by single stranded flexible hinge. The performance of DNA tweezers is dependent on repeated cycles. In each cycle the tweezers are either in open configuration or in closed configuration. These two configurations are interchanged in each cycleby adding two specific single stranded 'fuel' DNA sequences successively. A complete double stranded DNA sequence is produced in each cycle as the by-product.

A cycle of biochemical reactions of DNA tweezers is shown in Fig. 1 [Ray and Mondal, 2012]. Let, the cycle starts when the DNA tweezers are in open form. The domains *a*, *b*, *c*, *d* are marked in the figure. The complementary domains of *a* and *c* are marked as *a\** and *c\** respectively. The specific single stranded DNA sequence, input *A*, which works as fuel is added. Input *A* hybridizes to the two distal single stranded domains of the tweezers because of the Watson-Crick complementarity. The domains are *b* and *d*. The hybridization leads to the formation of closed configuration of the tweezers. Again another single stranded DNA sequence, input *B*, which works as fuel is added. Input *B* displaces the previously attached tweezers by branch migration and hybridize to input *A*. Again the open configuration of the tweezers is formed. The branch migration reaction occurs using domain *e* of input *B*as the toehold domain. The entire cycle will be repeated again and again as long as the 'fuel' sequences, i.e. input *A* and input *B*, are available. Each cycle generates a completely double stranded DNA sequence, termed as *by-product*.

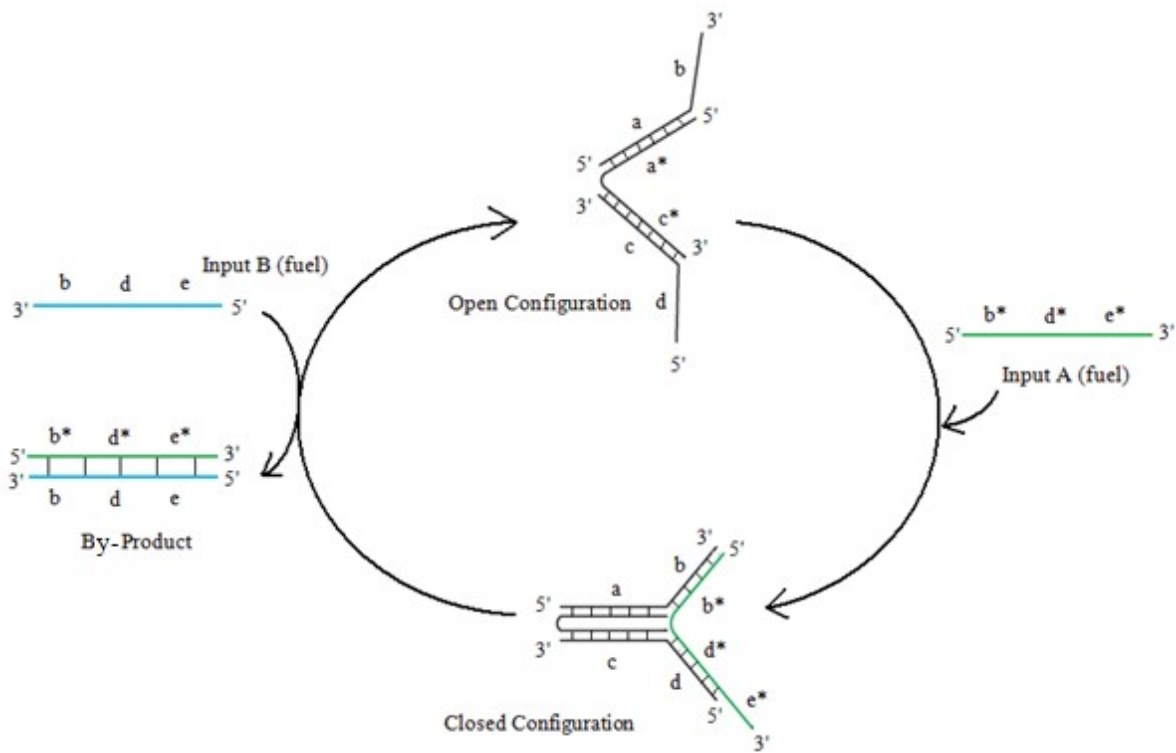

Figure 1.A complete cycle of DNA tweezers

For determination of the state of the device Yurke *et al.* [Yurke and Mills, 2003] used dye quenching procedure. The quencher molecule decreases the fluorescence intensity of the dye. The quenching efficiency is inversely proportional to the distance between the dye and the quencher molecule. For DNA tweezers, TET (5' tetrachloro-fluorescein phosphoramidite) is used as dye and TAMRA (carboxytetramethylrhodamine) is used as quencher. Thus, the greater fluorescence intensity indicates that the DNA tweezers are in open configuration and less intensity of the fluorescent dye indicates the closed configuration of the tweezers.

## 3. Syllogistic Reasoning

The Greek philosopher Aristotle (384 BC–322 BC) first proposed *syllogism*, an integral component of the formal study of logic, in his memorable piece of work on deductive reasoning, *Prior Analytics* (350 BC). Syllogism is a logical inference mechanism which generates conclusion based on two or more propositions, i.e. premises, by deductive reasoning. The syllogistic reasoning generally contains three propositions which are major premise, minor premise and conclusion. The famous example of syllogism developed by Aristotle is given below;

*Premise 1:* All men are mortal.
*Premise 2:* Socrates is a man.
*Conclusion:* Therefore, Socrates is mortal.

In this example premise 1 is the major premise and premise 2 is the minor premise. By testing minor premise against the major premise the plausible conclusion is drawn from the dispositional or propositional premises by deductive reasoning.

If the conclusion is deducted based on a chain of conditional statements i.e. premises, it is called *chaining syllogism*. The general form of chaining syllogism is shown below, where $p_1, p_2, \ldots \ldots p_n$ are the dispositions or propositions and $p_{n+1}$ is the deduced conclusion. If the premises are dispositions, the syllogistic reasoning will be considered under the domain of fuzzy logic. If the premises are propositions, it is considered as classical logic. In this paper we have considered propositions to perform syllogism.

$$p_1: \quad A_1 \text{ is } A_2$$
$$p_2: \quad A_2 \text{ is } A_3$$
$$.$$
$$.$$
$$.$$
$$\underline{p_n: \quad A_n \text{ is } A_{n+1}}$$

$$p_{n+1}: \quad A_1 \text{ is } A_{n+1}$$

If $p_{n+1}$ is deducted from $p_1, p_2, \ldots \ldots p_n$ then the chaining syllogism holds.

In section 5 we will formulate chaining syllogism by DNA tweezers [Ray and Mondal, 2012]. DNA tweezers are the DNA fueled device based on the mechanism of toehold mediated DNA strand displacement. Thus, in this paper the logical aspect of syllogism is replaced by DNA chemistry. The architecture of the DNA tweezers model is formally represented by process calculus in section 6 and graphically represented by DNA strand graph in section 7. In the next section we will briefly discuss the syntax and semantics of process calculus and DNA strand graph [Petersen *et. al.*, 2016].

## 4. Syntax and Semantics of Process Calculus and Strand Graph [Petersen *et. al.*, 2016, Ray and Mondal, 2017]

Petersen, Lakin and Phillips redefined DSD language and proposed a reformulated language, termed as *process calculus* [Petersen *et. al.*, 2016], to formulate the architecture of DNA models based on the mechanism of strand displacements in DNA strands with rich secondary structures (such as, branches and loops). In section 4.1 we will briefly describe the syntax and semantics of process calculus to formally model, simulate and analyze the complex, concurrent and communicating processes of DNA computation [Ray and Mondal, 2017].

### *4.1. Syntax and Semantics of Process Calculus*

In process calculus, a process or program $P$ is defined as a multiset of DNA strands $<S>$.

Process or program $P ::= <S_1> | ... | <S_i>$     where, $i \geq 0$

Each strand $<S>$ contains one or more domains $d$. Domain is actually a sequence of DNA bases or nucleotides i.e. A, T, G, C.

Strand $S ::= d_1 ..... d_i$     where, $i \geq 0$

A domain $d$ in a DNA strand is either free or bound with the complementary domain of any other DNA strand or to the same strand. A free domain is denoted by $d$. If the domain is bound by bond $x$, the bound domain is denoted by $d!x$. Let, an arbitrary domain is named $r$, then $r*$ is the complementary domain to which $r$ can bind by Watson-Crick base pairing. A domain is called *toehold* $t^\wedge$ if it is short enough to spontaneously unbind from its complement $t^\wedge*$.

The semantics of process calculus depends on some functions which determine whether a rule can be applied on a program. The functions are listed below;

- The function *comp*($r$) returns the complementary domain of domain $r$. Thus, it can be said that, *comp*($r$) = $r*$ and *comp*($r*$) = $r$.
- The function *toehold*($r$) returns true if $r$ is a toehold domain. Then we can also represent domain $r$ by $r^\wedge$.
- The function *adjacent(x, P)* returns the set of bonds that are adjacent to bond $x$ in program $P$.
- The function *hidden(x, P)* returns true if one end of bond $x$ occurs within a closed loop. Thus, the specific domain cannot bind to its complementary sequence.
- The function *anchored(x, P)* returns true if both ends of bond $x$ are held "close" to each other. Thus, bond $x$ is a part of a stable junction.
- The *context* $C(S_1, ..., S_i)$ is defined as a process $P$ containing sequences $S_1, ..., S_i$.
- The function *permute($S_1, ..., S_i$)* returns any possible permutation of sequences $S_1, ..., S_i$.

### *4.1.1. Semantics of Reduction rules of Process Calculus*

Now, we will define the semantics of some rules of process calculus by the following figures and corresponding expression.

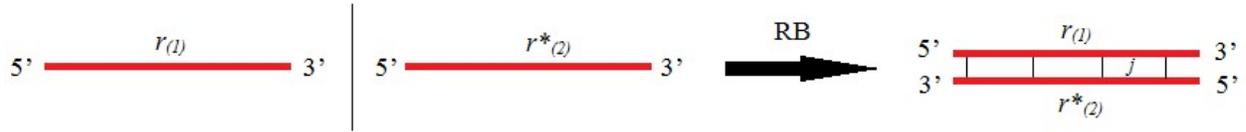

Figure 2. Rule (RB)

The semantics of rule (RB) as shown in Fig. 2 can be presented as,

$$(RB) \quad \frac{\neg hidden(j, P)}{C(r, r^*) \xrightarrow{RB, \{j\}} C(r!j, r^*!j) = P}$$

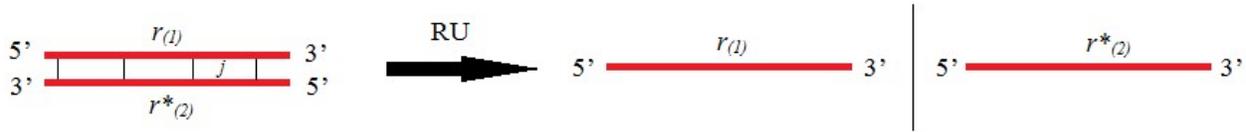

Figure 3. Rule (RU)

The semantics of rule (RU) as shown in Fig. 3 can be presented as,

$$(RU) \quad \frac{\neg anchored(j, P) \quad toehold(r)}{P = C(r!j, r^*!j) \xrightarrow{RU, \{i\}} C(r, r^*)}$$

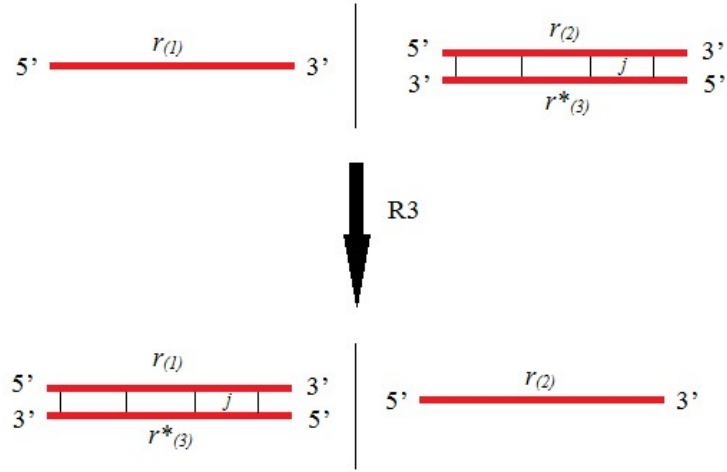

Figure 4. Rule (R3)

The semantics of rule (R3) as shown in Fig. 4 can be presented as,

$$(RU) \quad \frac{anchored(j, P)}{C(r, r!j, r^*!j) \xrightarrow{R3,\{j\}} C(r!j, r, r^*!j) = P}$$

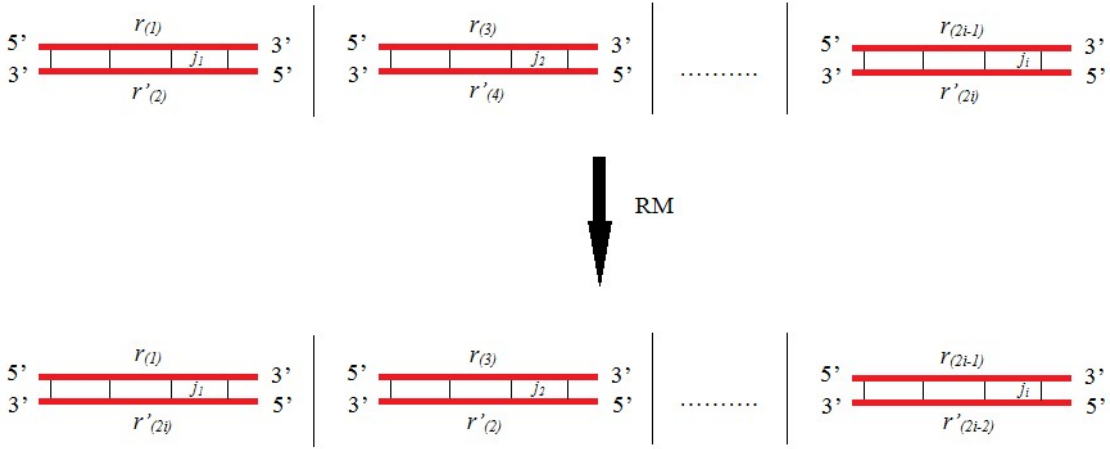

Figure 5. Rule (RM)

The semantics of rule (RM) as shown in Fig. 5 can be presented as,

$$(RM) \quad \frac{anchored(j_1, P') \ldots \ldots anchored(j_i, P')}{C(r!j_1, r'!j_1, r!j_2, r'!j_2, \ldots .r!j_i, r'!j_i) \xrightarrow{RM,\{j_1 \ldots j_i\}} C(r!j_1, r'!j_2, r!j_2, r'!j_3, \ldots .r!j_i, r'!j_1) = P'}$$

The reduction rules of process calculus are described by the help of an example in our paper [Ray and Mondal, 2017]. In that example we have given program codes for hairpin toehold exchange mechanism with two invader strands.

But there are some limitations of process calculus in implementation of different rules, because of the complexity of pattern matching on arbitrary process contexts. To overcome this problem Petersen *et. al.* [Petersen et. al., 2016] introduces the concept of DNA strand graph. The syntax and semantics of strand graph is briefly described in the next section [Ray and Mondal, 2017].

## *4.2. Syntax and Semantics of DNA Strand Graph [Petersen et. al., 2016, Ray and Mondal, 2017]*

Graphs are mathematical structures which are used to model pair-wise relations between objects. The graphical structures are formed by vertices or nodes which are connected by edges. In a graph if there is no distinction between the two nodes associated with each edge, the graph is said to be undirected. In directed graph each edge has a specific direction from one node to another. In strand graph the expressive power of graph theory can represent rich secondary structures of DNA strands and implement the complex rules. Now we will summarize the notation for strand graph theory as demonstrated in the paper [Petersen *et. al.*, 2016].

Strand graph is defined by $G = (V, length, colour, A, toehold, E)$, where,

$V = \{1,......,N\}$ denotes the *set of vertices* of the graph. Each vertex, shown by natural number, represents a *DNA strand*. There are different *sites* in a vertex. Each *site s* denotes a specific *domain* of that strand. The vertices are drawn as circular arrow with a specific direction i.e. from 5' to 3' of a DNA strand. The sites are placed in a vertex according to the occurrences of the corresponding domain in the specific strand. Site is represented as $s = (s, n)$, where $v$ is a vertex and $n$ is the position of site $s$ in vertex $v$. Both $v$ and $n$ are natural numbers.

*length:* denotes a *function* which assigns a *specific length to each vertex*. Lengths are represented by natural numbers.

*colour:* denotes a *function* which assigns a *specific colour to each vertex*. Colours are also represented by natural numbers. Thus, it would be easier to identify a particular vertex representing a specific DNA strand. Colour is actually a function of the length. If $v_1$ and $v_2$ are two vertices of a strand graph, then, $length(v_1) = length(v_2) \Rightarrow colour(v_1) = colour(v_2)$.

$A$ is the set of *admissible edges* of the strand graph. If two domains of the DNA strands are complementary, they are able to hybridize with each other by forming a bond. Then an edge can be drawn between the sites of the vertices representing those domains. Throughout the performance of the whole program, all bonds those are allowed to be formed are represented by the set of admissible edges. Edge is represented as $e = \{s_1, s_2\}$ where $s_1$ and $s_2$ are two sites and $s_1 \neq s_2$. Again, we can write that, $e = \{(v_1, n_1), (v_2, n_2)\}$.

*Toehold* is a function that returns true if admissible edges exist between the short domains i.e. toehold domains and returns false for admissible edges between the long domains.

$E$ is the set of current edges of the strand graph which is expressed as $\{e_1, .....,e_l\} \subseteq A$. In the contrary of other above mentioned information to define strand graph, $E$ is non-static information. During the execution of the program the set of current edges changes with the change in reduction rules. A domain in a DNA strand cannot bind with more than one domain at any given instant i.e. only one edge can be drawn from a given site at that point of time. This is can be expressed as, $(i \neq j) \Rightarrow e_i \cap e_j = \emptyset$.

The following definition can be written to define a DNA strand graph [Petersen *et. al.*, 2016] using the above explained function;

$V \quad\quad\quad\quad\quad\quad = \quad \{1, ....., N\} \quad\quad\quad$ where, $N$ is natural number
$length(v) \quad\quad\quad = \quad len(Sv)$
$colour(v) = i \quad\quad \Leftrightarrow \quad tp(Sv) = t_i$
$(v_1, n_1) \stackrel{A}{\leftrightarrow} (v_2, n_2) \quad \Leftrightarrow \quad ndom(v_1, n_1) = comp(ndom(v_2, n_2))$
$toehold(\{s_1, s_2\}) \quad \Leftrightarrow \quad toe(s_1)$
$(v_1, n_1) \stackrel{E}{\leftrightarrow} (v_2, n_2) \quad \Leftrightarrow \quad \exists d, j. dom(v_1, n_1) = d!j \land dom(v_2, n_2) = comp(d)!j$

where, $d$ denotes the domain and $j$ denotes the bond between $(v_1, n_1)$ and $(v_2, n_2)$.

In the next section we will illustrate the semantics of reduction rules.

*4.2.1. Semantics of reduction rules of DNA strand graph*

DNA strand graph transits from one state to another by following the reduction rules. The change in state of the strand graph is indicated by the change in colours of the edges among vertices. The semantics of the reduction rules need definitions of few functions [Petersen *et. al.*, 2016].

The function *sites(E)* returns the set of sites in set of current edges $E$ which can be expressed by $\{s | \exists e \in E. s \in e\}$.

If two edges in a strand graph not only exist between the same pair of vertices but also the corresponding sites are adjacent to each other, the two edges are said to be adjacent. The function *adjacent(e, E)* returns the set of adjacent edges to edge $e$ from the set $E$.

The function *hidden(e, E)* returns true if one of the ends of edge $e$ from the set $E$ occurs within a closed loop.

The function *anchored(e, E)* returns true if the edge $e$ from the set $E$ is a part of a stable junction by holding the corresponding sites close to each other.

Now we will describe the semantics of reduction rules [Petersen *et. al.*, 2016] through which the program occurs and reaches to its final state.

**Rule (GB)**

Let the sites of two vertices of a DNA strand graph is joined by admissible edge $x$ which is not current at that instant. If those two sites are not preoccupied and open to each other, according to *rule (GB)* $x$ can be converted into current edge. The semantics of rule (GB) is given below;

$$(GB) \frac{x \in A \backslash E \quad x \cap sites(E) = \emptyset \quad \neg hidden(x, E)}{E \xrightarrow{GB, \{x\}} E \cup \{x\}}$$

**Rule (GU)**

Let the sites of two vertices of a DNA strand graph is joined by admissible edge $e$ and the sites represent toehold domain. Toehold domains are short enough to spontaneously unbind from its complement. Thus according to *rule (GU)* if the toehold domains are not anchored, the edge $e$ can be removed from the current set $E$ of the corresponding strand graph. The semantics of rule (GU) is given below;

$$(GU) \frac{e \in E \quad toehold(e) \quad \neg anchored(e, E)}{E \xrightarrow{GU, \{e\}} E \backslash \{e\}}$$

**Rule (G3)**

Let the sites of two vertices of a DNA strand graph is joined by admissible edge $x$ which is not current at that instant. $x$ can be joined to the set of current edges $E$ even though one of the end sites is preoccupied by some other site forming a current edge $e$. $x$ becomes current edge by removing $e$ if the function *anchored(x, E)* returns true. This mechanism is termed as *displacing path*. The swapping of single bonds can form a long chain through the whole program. This mechanism is performed by reduction *rule (G3)*. The semantics of rule (G3) is given below;

$$(G3) \frac{e \in E \quad x \in A\backslash E \quad e = \{s, s'\} \quad x = \{s, s''\} s'' \notin sites(E) \quad anchored(x, E)}{E \xrightarrow{G3,\{x\}} (E\{e\}) \cup \{x\}}$$

*Rule (GM)*

By the reduction *rule (GM)* the mechanism of displacing path i.e. swapping of single bonds makes a loop. The semantics of rule (GM) is given below;

$$(GM) \frac{i \in \{1,\ldots,N\} e_i \in E \quad x_i \in A\backslash E \quad e_i = \{s_i, s'_i\} x_i = \{s'_{i-1}, s_i\} s'_0 = s'_N \quad anchored(x_i, E)}{E \xrightarrow{GM,\{x_1,\ldots,x_N\}} (E\backslash\{e_1,\ldots,e_N\}) \cup \{x_1,\ldots,x_N\}}$$

DNA strand graph and the reduction rules have been explained by the example of toehold-mediated four-way strand displacement and branch migration in our paper [Ray and Mondal, 2017].

In the next section we will formulate chaining syllogism by DNA tweezers [Ray and Mondal, 2012].

## 5. Formulation of chaining syllogism by DNA tweezers [Ray and Mondal, 2012]

Let us consider, we are given with three premises (i.e. $p_1, p_2, p_3$) which are propositions. Each of the propositions has two clause; the first one is the antecedent clause and the second one is the consequent clause.

$$\left.\begin{array}{ll} p_1: & \textit{icy roads are slippey} \\ p_2: & \textit{slippery roads are risky} \\ p_3: & \textit{risky roads are accident prone} \end{array}\right\} S$$

We will have to deduct the plausible conclusion of the above set of propositions ($S$) by chaining syllogistic reasoning. We have replaced the logical aspect of the syllogistic reasoning by the biochemical operations which can manipulate DNA strands. In other words we can say that, in this section we will perform logical reasoning by DNA strands using the dynamic molecular device, DNA tweezers. The set of propositions $S$ can be represented by DNA tweezers which is the ontological representation of icy roads and its adverse consequences.

There are four domains (icy, slippery, risky and accident prone) in the given premises. Each of the domain is encoded by arbitrarily chosen ten bases long DNA oligonucleotide or its complementary sequence. The list of encoded DNA oligonucleotides and their abbreviations are given in Table 1.

| Domain | Abbreviation | Encoded DNA strand |
|---|---|---|
| Icy | $i$ | 5′ − CATGCTAGGC − 3′ |
|  | $i^*$ | 3′ − GTACGATCCG − 5′ |
| Slippery | $s$ | 5′ − TGCAGCCAAT − 3′ |
|  | $s^*$ | 3′ − ACGTCGGTTA − 5′ |
| Risky | $r$ | 5′ − AGTGCACTGC − 3′ |
|  | $r^*$ | 3′ − TCACGTGACG − 5′ |
| Accident prone | $a$ | 5′ − GCTGACTCGA − 3′ |
|  | $a^*$ | 3′ − CGACTGAGCT − 5′ |

Table 1. Representation of the domains by DNA strands

Now we will encode the set of given premises $S$ by DNA sequences to construct the DNA tweezers.

$p_1$: ***icy roads are slippey***

The premise or proposition $p_1$ consists of two domains. To code the proposition, the domains are encoded by the corresponding DNA sequences as demonstrated in Table 1 in 5' to 3' direction.

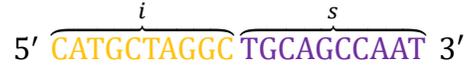

$p_2$: ***slippery roads are risky***

The premise or proposition $p_2$ also consists of two domains; slippery and risky. This proposition is encoded by the DNA strand in 3' to 5' direction (Table 1). A four bases long arbitrarily chosen DNA sequence, termed as spacer, is incorporated in between two domains, slippery and risky. We have chosen ATGC (in 5' to 3' direction) as the spacer ($sp$).

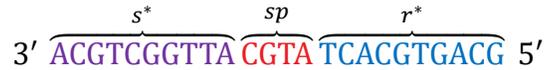

$p_3$: ***riskyroadsareaccidentprone***

The proposition $p_3$ consists of two domains. The domains, risky and accident prone, are encoded by the corresponding DNA sequences as given in Table 1 in 5' to 3' direction.

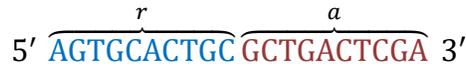

The 5' and 3' ends of the DNA strand encoding $p_2$ are labeled with dyes TET (5' tetrachloro-fluorescein phosphoramidite) and TAMRA (carboxytetramethylrhodamine), respectively.

Thus, all the premises of $S$ are encoded in form of single stranded DNA sequences. Using the above three strands DNA tweezers in open configuration can be constructed (Fig. 6).

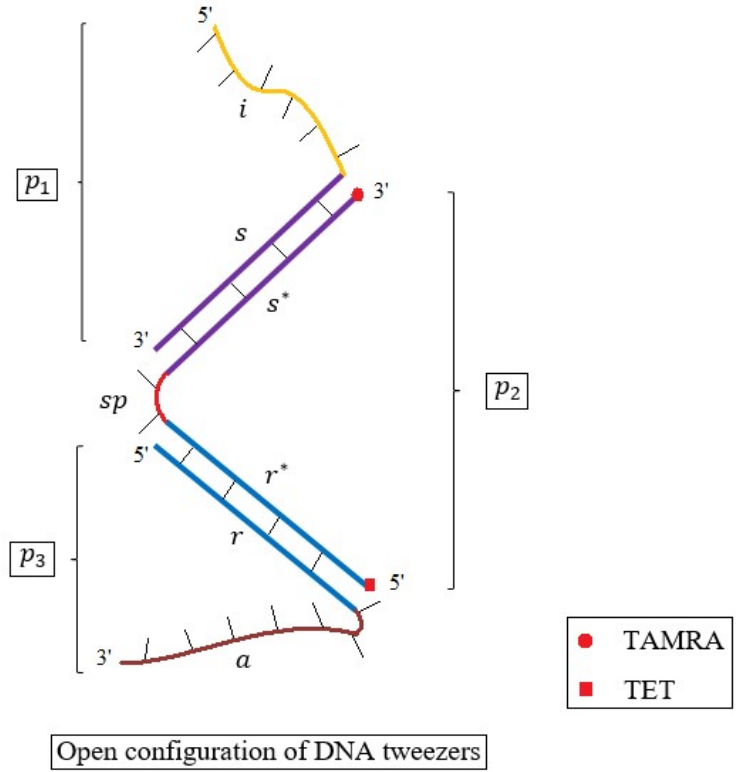

Figure 6. Representation of a set of premises by DNA tweezers

From the given set of propositions or premises $S$ we have to deduct the plausible conclusion by syllogistic reasoning. We are going to perform reasoning by DNA computing using the encoded DNA strands. We will follow the backward chaining procedure which proceeds from a tentative conclusion backward to the premise to determine if the given data supports that conclusion [Ray and Mondal, 2012]. Finally the specific DNA strand which encodes the possible conclusion has to be separated. In the next subsection 5.1, we will describe the wet lab algorithm for syllogistic reasoning using DNA tweezers.

### 5.1. Wet lab algorithm
*Step 1.*
All the premises of $S$ are encoded by single stranded DNA oligonucleotide as described above (section 5). These three encoded strands are hybridized with each other to form the open configuration of DNA tweezers (Fig. 6).

*Step 2.*
We have to deduct the plausible conclusion of the given set of premises. In our algorithm backward chaining procedure is followed which proceeds from a tentative conclusion backward to the premise to determine if the given data supports that conclusion. A database of possible conclusions is formed containing different single stranded DNA sequences in 3' to 5' direction.

The possible conclusions of the given propositions are predicted. These are called the hypotheses. The hypotheses of this particular reasoning problem can be; icy roads are risky, risky roads are slippery, accident prone roads are slippery, icy roads are accident prone, slippery roads are icy etc. At the 5' end of each encoded hypothesis (in 3' to 5' direction) another five bases long oligonucleotide, GGCAT, is attached. Later, it will act as the complementary to toehold domain in DNA strand displacement. This domain is abbreviated as $t^{\wedge *}$. These single stranded sequences, termed as input *A*, are added in the test tube containing the DNA tweezers.

Some of the hypotheses are shown in Fig. 7.

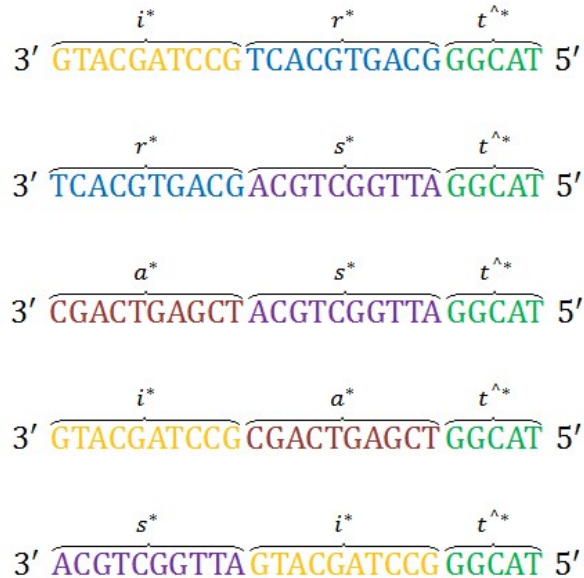

Figure 7. Input *A* - DNA strands representing hypotheses

*Step 3.*

Only one strands from the above mentioned set of hypotheses, i.e. input *A*, gets hybridized to the open configuration of the DNA tweezers (Fig. 6). That particular single stranded DNA sequence is completely complementary to single stranded part of the DNA tweezers and leads to the formation of closed configuration.

Input *A* works backward from consequent to the antecedent to see if any of the sequences can completely hybridize to the partially double stranded tweezers.

In closed configuration of the tweezers the fluorescence intensity decreases as the attached dye molecules (TET) and quencher molecules (TAMRA) come closer.

*Step 4.*

Another set of single stranded DNA sequences, termed as input *B*, is added in the solution containing DNA tweezers and input *A*. The sequences added as input *B* are the complementary strands of input *A*. Some of these sequences are shown in Fig. 8.

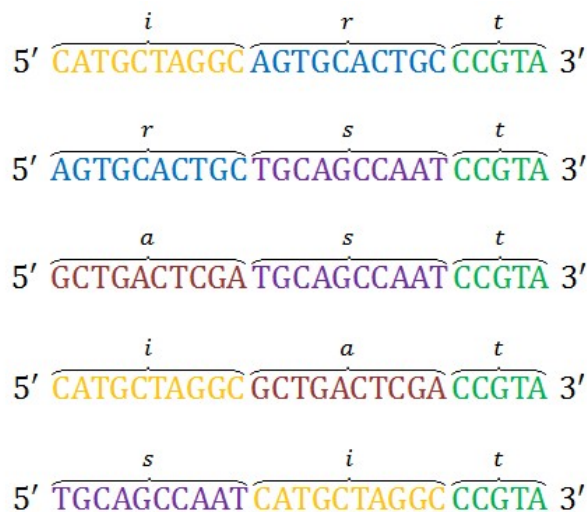

Figure 8. Input *B* - Complementary to Input *A*

*Step 5.*

From the sequences added as input *B*, only one sequence has the ability to displace previously attached input *A* from the closed complex. Here, the toehold mediated branch migration and strand displacement occurs. The five bases of the 3' end of input B is the toehold domain. This domain first hybridizes with the hanging single stranded complementary domain of the closed complex and gradually hybridization progress through brand migration.

One cycle of reactions completes when input *B* completely hybridizes to input *A* by displacing the partially double stranded DNA tweezers. The completion of each cycle leads to the formation of a complete double stranded DNA by-product. The closed configuration of the tweezers again returns to its open form. The entire mechanism of DNA tweezers and the formation of double stranded by-product is shown in Fig. 9.

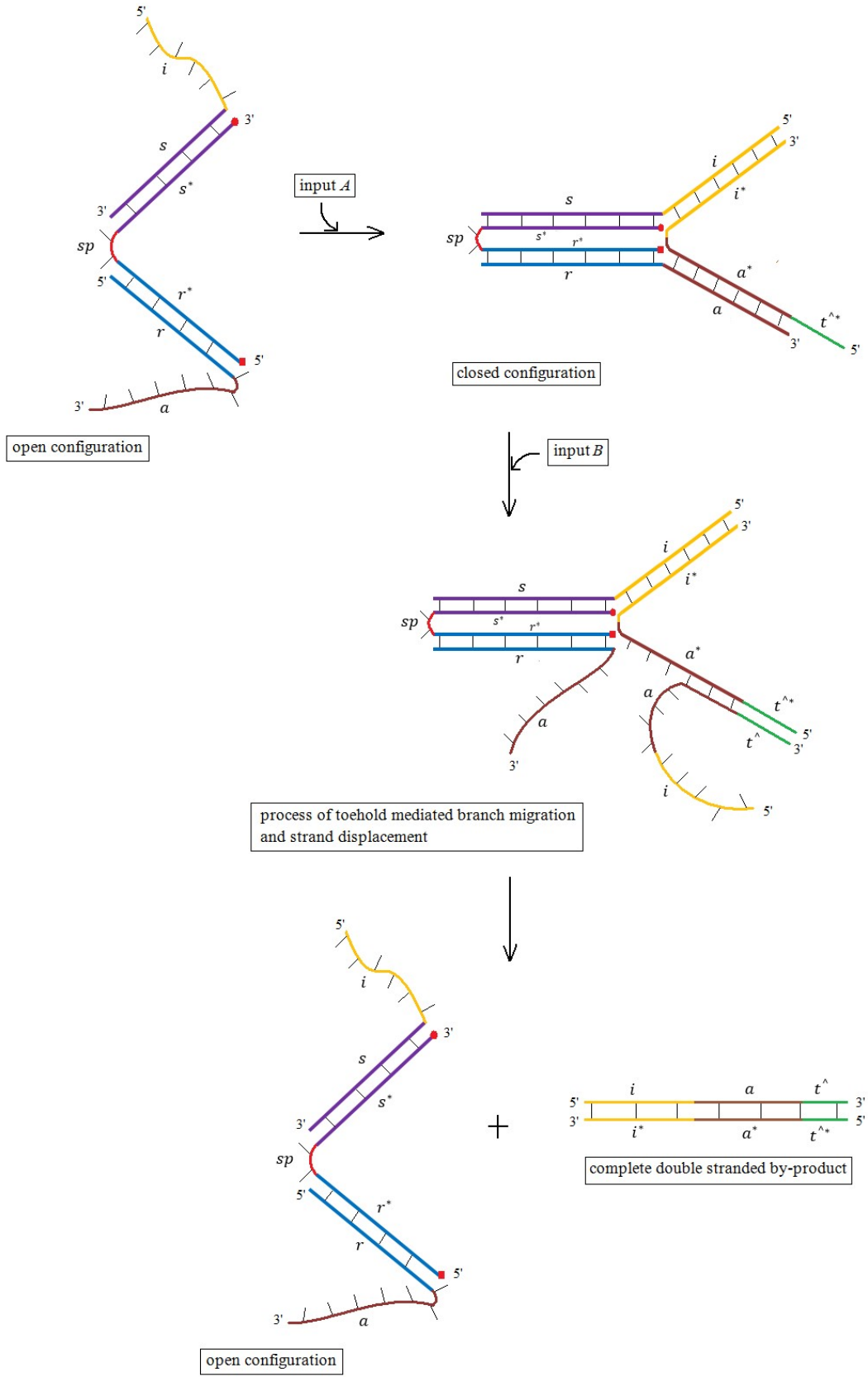

Figure 9. Mechanism to perform syllogistic reasoning by DNA tweezers

*Step 6.*

After adding input *A* and input *B* respectively to the initial solution, it is divided in two sample test tubes. The first sample tube is treated with exonuclease enzyme and the other test tube is kept untreated. The partial double stranded sequences or the strands with sticky ends in the first sample deteriorates because of the addition of exonuclease. Complete double stranded DNA sequences remains unaffected.

*Step 7.*

Gel electrophoresis is performed with the DNA samples of two test tubes. Final conclusion can be drawn by comparing the electrophoretograms for these samples. The electrograms show the existence or non-existence of the desired double stranded molecule in the reaction test tube. If the DNA device performs syllogistic reasoning and gives a precise conclusion, the location of at least one DNA band remains unchanged on both of the electrophoretograms. This band is formed because of the by-product of the reaction cycle which is completely hybridized double stranded DNA sequence without any sticky end.

*Step 8.*

The by-product of the algorithm is the desirable conclusion derived from the given set of proposition *S*. The order of the bases of the DNA strand encoding the plausible conclusion, can be known from sequencer. The by-product contains three domains. The first domain represents the antecedent clause, the second domain is the consequent clause and third one is toehold required for strand displacement.

By decoding the sequence representing the by-product according to Table 1, we can conclude that the conclusion of *S* by performing syllogistic reasoning using DNA tweezers is, "*Icy roads are accident prone*".

In the next section we will code syllogistic reasoning by DNA tweezers by the formal language, process calculus, using the syntax and semantics described in subsection 4.1.

## 6. Syllogistic reasoning by DNA tweezers coded using process calculus

Let the *P* is the program which performs syllogistic reasoning by DNA tweezers. The program *P* consists of five DNA strands. Three of these strands ($S_1$, $S_2$ and $S_3$) codes the set of premises (*S*) i.e. $p_1$, $p_2$ and $p_3$. The remaining two strands, $S_4$ and $S_5$, code input *A* and input *B* respectively. Thus, *P* can be defined as the multiset of five DNA strands.

$$P = <S_1> \mid <S_2> \mid <S_3> \mid <S_4> \mid <S_5>$$

Therefore, *P* can be written as,

$$P = <i \quad s!x> \mid <r^*!y \quad sp \quad s^*!x> \mid <r!y \quad a> \mid <t^{\wedge *} \quad a^* \quad i^*> \mid <i \quad a \quad t^{\wedge}>$$

where, all the strands are shown in 5' to 3' direction.

The literals are encoded by arbitrarily chosen ten bases long single-stranded DNA sequence representing the domains of the corresponding DNA strands (see Table 1). But, the toehold domain, i.e. $t^\wedge$ and $t^{\wedge*}$, is five bases long DNA oligonucleotide. It is short enough to spontaneously hybridize and unhybridize to its complementary sequence. The given program code shows that domain *s* at the 3' end of <$S_1$> is bound to domain *s\** of <$S_2$> by bond *x*. The domain *r\** of <$S_2$> is bound to domain *r* of the DNA strand <$S_3$> by bond *y*. Thus, the DNA tweezers is formed by partially hybridized strands <$S_1$>, <$S_2$> and <$S_3$>. At this stage the tweezers are in open form. <$S_4$> codes input *A* and <$S_4$> codes input *B*. Initially, <$S_4$> and <$S_5$> are single stranded as all the domains of the corresponding strands are free.

As the domain *i* of <$S_1$> and the domain *i\** of <$S_4$> are not bound with any other domain, the program matches the context $C(i, i^*)$. It can be written that $P' = C(i!u_1, i^*!u_1)$ as one end of the bond $u_1$ is not in closed loop, i.e. $hidden(u_1, P)$ returns false. Thus, the program $P'$ can be produced by the *rule (RB)* which forms the new bond $u_1$ between the complementary domains *i* and *i\**. The program code is given below;

$<\!i\ \ s!x\!>\ |\ <\!r^*!y\ \ sp\ \ s^*!x\!>\ |\ <\!r!y\ \ a\!>\ |\ <\!t^{\wedge*}\ \ a^*\ \ i^*\!>\ |\ <\!i\ \ a\ \ t^\wedge\!>$
$\stackrel{(RB)}{\longrightarrow} <\!i!u_1\ \ s!x\!>\ |\ <\!r^*!y\ \ sp\ \ s^*!x\!>\ |\ <\!r!y\ \ a\!>\ |\ <\!t^{\wedge*}\ \ a^*\ \ i^*!u_1\!>\ |\ <\!i\ \ a\ \ t^\wedge\!>$

Again, the domain *a* of <$S_3$> and the domain *a\** of <$S_4$> are free. Thus, the program matches the context $C(a, a^*)$. New bond $v_1$ can be formed by the *rule (RB)* between the complementary domains *a* and *a\**. The formation of bond $v_1$ is possible, if and only if the program $hidden(v_1, P)$ returns false. This step leads the formation of closed configuration of the DNA tweezers. The program code is given below;

$<\!i!u_1\ \ s!x\!>\ |\ <\!r^*!y\ \ sp\ \ s^*!x\!>\ |\ <\!r!y\ \ a\!>\ |\ <\!t^{\wedge*}\ \ a^*\ \ i^*!u_1\!>\ |\ <\!i\ \ a\ \ t^\wedge\!>$
$\stackrel{(RB)}{\longrightarrow} <\!i!u_1\ \ s!x\!>\ |\ <\!r^*!y\ \ sp\ \ s^*!x\!>\ |\ <\!r!y\ \ a!v_1\!>\ |\ <\!t^{\wedge*}\ \ a^*!v_1\ \ i^*!u_1\!>\ |\ <\!i\ \ a\ \ t^\wedge\!>$

Now, <$S_5$> i.e. input *B* comes into action. The function *toehold(t)* returns true for <$S_5$>. The toehold domain $t^\wedge$ at the 3' end of <$S_5$> has a free complementary domain $t^{\wedge*}$ in <$S_4$>. The program matches the context $C(t^\wedge, t^{\wedge*})$. Thus, it can be written that $P' = C(t^\wedge!w, t^{\wedge*}!w)$ as one end of the bond *w* is not in closed loop, i.e. $hidden(w, P)$ returns false. Thus, according to *rule (RB)* the program $P'$ can be generated which forms the new bond *w* between the toehold and its complementary domain. The program code is shown below;

$<\!i!u_1\ \ s!x\!>\ |\ <\!r^*!y\ \ sp\ \ s^*!x\!>\ |\ <\!r!y\ \ a!v_1\!>\ |\ <\!t^{\wedge*}\ \ a^*!v_1\ \ i^*!u_1\!>\ |\ <\!i\ \ a\ \ t^\wedge\!>$
$\stackrel{(RB)}{\longrightarrow} <\!i!u_1\ \ s!x\!>\ |\ <\!r^*!y\ \ sp\ \ s^*!x\!>\ |\ <\!r!y\ \ a!v_1\!>\ |\ <\!t^{\wedge*}!w\ \ a^*!v_1\ \ i^*!u_1\!>\ |\ <\!i\ \ a\ \ t^\wedge!w\!>$

The toehold domain $t^\wedge$ is short enough to unbind spontaneously. As the newly formed bond *w* is not a part of a junction that holds both ends of the bond close to each other, the program *anchored(w, P)* returns false. Thus, according to *rule (RU)* the bond *w* between the toehold and its complementary domain can be broken to generate the program $C(t^\wedge, t^{\wedge*})$. It is reversible of rule (RB). The program code is shown below;

$<i!u_1 \quad s!x> | <r^*!y \quad sp \quad s^*!x > | <r!y \quad a!v_1> | <t^{\wedge *}!w \quad a^*!v_1 \quad i^*!u_1> | <i \quad a \quad t^{\wedge}!w >$
$\xrightarrow{(RU)} <i!u_1 \quad s!x> | <r^*!y \quad sp \quad s^*!x > | <r!y \quad a!v_1> | <t^{\wedge *} \quad a^*!v_1 \quad i^*!u_1> | <i \quad a \quad t^{\wedge}>$

Because of the hybridization of toehold domains, toehold mediated branch migration and strand displacement occurs. By strand displacement two DNA strands with partial or full complementarity hybridize to each other, displacing one or more pre-hybridized strands [Zhang and Seelig, 2011]. The free domain $a$ of $<S_5>$ has a complementary domain $a^*$ in $<S_4>$ which is already bound by the bond $v_1$. In this step the program matches the context $C(a!v_1, a^*!v_1, a)$. It should be checked that, if an anchored bond can be formed between domains mentioned above to generate the program $P' = C(a, a^*!v_2, a^*!v_2)$. In this step, a new bond $v_2$ can be generated by applying *rule (R3)* as there is a bond $w$ that is immediately adjacent to $v_2$ in $P'$, holding both ends of bond $v_2$ close to each other.

$<i!u_1 \quad s!x> | <r^*!y \quad sp \quad s^*!x > | <r!y \quad \boldsymbol{a!v_1}> | <t^{\wedge *}!w \quad \boldsymbol{a^*!v_1} \quad i^*!u_1> | <i \quad \boldsymbol{a} \quad t^{\wedge}!w >$
$\xrightarrow{(R3)} <i!u_1 \quad s!x> | <r^*!y \quad sp \quad s^*!x > | <r!y \quad \boldsymbol{a}> | <t^{\wedge *}!w \quad \boldsymbol{a^*!v_2} \quad i^*!u_1> | <i \quad \boldsymbol{a!v_2} \quad t^{\wedge}!w >$

In this step the branch migration and strand displacement continues to occur. $<S_5>$ displaces the DNA tweezers and completely hybridizes to $<S_4>$. The free domain $i$ of $<S_5>$ has a complementary domain $i^*$ in the strand $<S_4>$. But $i^*$ is preoccupied by domain $i$ of $<S_1>$ by forming the bond $u_1$. According to *rule (R3)* $u_1$ can be broken and a new bond $u_2$ can be generated between the domain $i$ of $<S_5>$ and its complementary domain in $<S_4>$, as there is a bond $v_2$ immediately adjacent it which holds both ends of bond $u_2$ close to each other.

$<\boldsymbol{i!u_1} \quad s!x> | <r^*!y \quad sp \quad s^*!x > | <r!y \quad a> | <t^{\wedge *}!w \quad a^*!v_2 \quad \boldsymbol{i^*!u_1}> | <\boldsymbol{i} \quad a!v_2 \quad t^{\wedge}!w >$
$\xrightarrow{(R3)} <\boldsymbol{i} \quad s!x> | <r^*!y \quad sp \quad s^*!x > | <r!y \quad a> | <t^{\wedge *}!w \quad a^*!v_2 \quad \boldsymbol{i^*!u_2}> | <\boldsymbol{i!u_2} \quad a!v_2 \quad t^{\wedge}!w >$

The resultant code of the program $P$ shows that, again the DNA tweezers return to its open configuration. The DNA strands $<S_4>$, i.e. input $A$, and $<S_5>$, i.e. input $B$, are completely bound to each other by three newly formed bonds $w$, $v_2$ and $u_2$. This complete double stranded DNA sequence is the by-product of the entire program coded above. The domains of the by-product encode the conclusion of the set of propositions $S$. The chaining syllogism has been solved using strand displacement mechanism of DNA tweezers and in this section the entire procedure is formally coded by process calculus.

## 7. Representation of syllogistic reasoning by DNA tweezers using DNA strand graph and reduction rules

In this section we will graphically represent the wet lab algorithm to perform syllogistic reasoning by DNA tweezers (section 5). By DNA strand graph, the architecture of tweezers model can be analyzed more expressively. In this paper we are deducing conclusion from a given set of proposition $S$. The reasoning aspect has been replaced by DNA chemistry which is coded by program $P$ in section 6. At the initial stage of the reaction the program code is expressed as,

$P = < i \quad s!x > | < r^*!y \quad sp \quad s^*!x > | < r!y \quad a > | < t^{\wedge *} \quad a^* \quad i^* > | < i \quad a \quad t^{\wedge} >$

The graphical depiction of program *P* is the DNA strand graph *G* shown in Fig. 10.

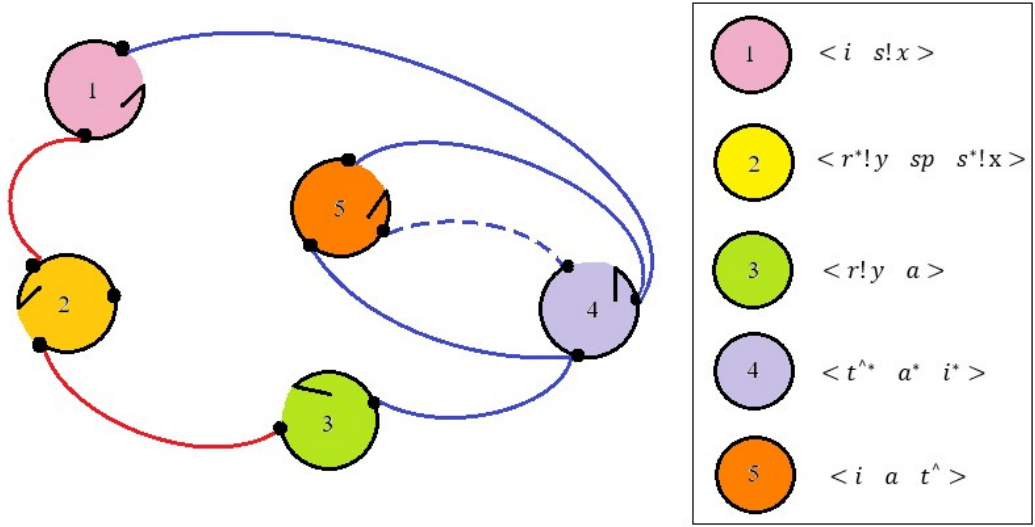

Figure 10. DNA strand graph *G* representing the initial state of program *P*

Each of the five strands in program *P* is represented by the vertices in graph *G* (Fig. 10). Arbitrary colours are assigned for the vertices in the graph. The vertices are drawn as circular arrows and the arrowhead indicates the 3' end of the DNA strand. The domains of the DNA strands are presented by the sites which are placed on the arrow-headed vertices according to their occurrences. All the admissible edges are drawn connecting the corresponding sites of the vertices. The current edges are represented by red lines and the remaining edges are represented by blue lines. The toehold edges are shown by dashed lines.

The initial state of DNA strand graph, shown in Fig. 10, is defined by *G = (V, length, colour, A, toehold, E)*, where,

| | | |
|---|---|---|
| *V* | = | {1, 2, 3, 4, 5}. |
| *length* | = | {1 → 2, 2 → 3, 3→ 2, 4 → 3, 5 → 3}. |
| *colour* | = | {1 → 1, 2 → 2, 3→ 3, 4 → 4, 5 → 5}. |
| *A* | = | {(1, 1), (4, 3)},{(1, 2), (2, 3)},{(2, 1), (3, 1)},{(3, 2), (4, 2)},{(4, 1), (5, 3)}, {(4, 2), (5, 2)}, {(4, 3), (5, 1)}. |
| *toehold* | = | {(4, 1), (5, 3)}. |
| *E* | = | {(1, 2), (2, 3)}, {(2, 1), (3, 1)}. |

Fig. 11 is the graphical representation of program *P* which performs syllogistic reasoning using DNA strands. Each step of the program changes according to suitable reduction rule.

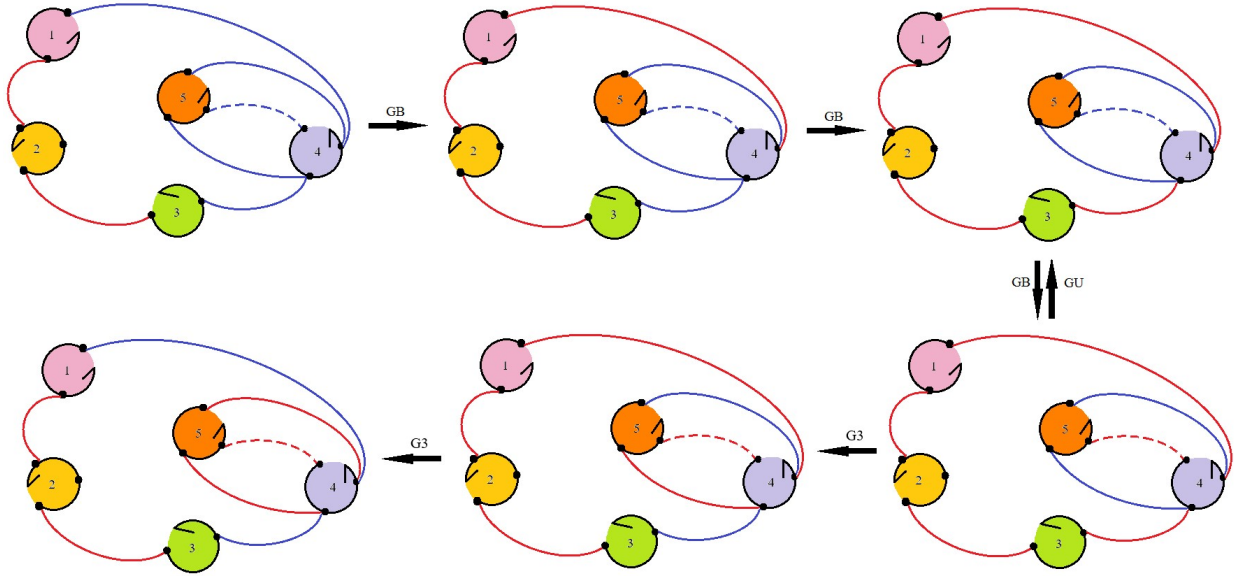

Figure 11. DNA strand graph with reduction rules representing the program *P* which performs syllogistic reasoning using DNA tweezers

Initially only the admissible edges connecting the sites of first three vertices are included in the set of current edges *E*. Thus, the DNA tweezers are in open configuration; and input *A* (represented by vertex 4) and input *B* (represented by vertex 5) are in single stranded form. The 1$^{st}$ site (domain *i*) of vertex 1 and the 3$^{rd}$ site (domain *i\**) of vertex 4 are not preoccupied and open to each other. Thus, according to *rule (GB)* in the first step of the program the admissible edge joining these two sites is converted to current edge. Thus, the colour of the edge changes from blue to red. Similarly, in the next step the admissible edge joining 2$^{nd}$ site (domain *a*) vertex 3 and the 2$^{nd}$ site (domain *a\**) of vertex 4 becomes current according to *reduction rule (GB)*. This step leads to the closed configuration of the DNA tweezers.

Now, the toehold domain *t^* (3$^{rd}$ site) of vertex 5 hybridizes to the complementary domain *t^\** (1$^{st}$ site) of vertex 4 and the blue dashed line is converted to red. This step is reversible. After the hybridization of the toehold domains, the branch migration and strand displacement occurs. The 2$^{nd}$ site (domain *a\**) of vertex 4 is preoccupied by the 2$^{nd}$ site (domain *a*) of vertex 3. The edge joining these two sites is omitted from the set of current edge and the admissible edge joining 2$^{nd}$ site (domain *a\**) of vertex 4 and 2$^{nd}$ site (domain *a*) of vertex 5 is included in the set of current edges. The strand displacement continues as the red edge joining the 1$^{st}$ site (domain *i*) of vertex 1 and the 3$^{rd}$ site (domain *i\**) of vertex 4 becomes blue; and the blue edge joining the 3$^{rd}$ site (domain *i\**) of vertex 4 and 1$^{st}$ site (domain *i*) of vertex 5 becomes blue. The strand displacement occurs by the *reduction rule (G3)*.

Now, all the admissible edges joining the corresponding sites of vertex 4 and vertex 5 are included in the set of current edges. Thus, it can be said that input *A* (vertex 4) is completely bound with input *B* (vertex 5). This complete double stranded DNA sequence is the resultant by-product of the program *P* and the conclusion of the chaining syllogism (the set of propositions, *S*, as stated in section 5). The DNA tweezers returns to its initial open configuration. The cycle is repeated again and again as long as the fuel molecules, input *A* and input *B*, is available.

## 8. Conclusion

In this paper we have illustrated how syllogistic reasoning by DNA tweezers can be presented by the semantics of process calculus and DNA strand graph. We have used the chemical potential and flexibility of DNA strands to generate a robust but simple DNA-fuelled dynamic device to perform syllogistic reasoning which is essential for commonsense reasoning of an individual. By using formal language theory in form of process calculus and the expressive power of DNA strand graph, we have successfully modeled, analyzed and simulated the molecular machine to perform syllogism from a given set of propositions. Propositions are considered as two-valued logic or classical logic.

This work can further be extended to perform reasoning with dispositions. The dispositions are basically propositions which are preponderantly but not necessarily always true [Zadeh, 1985]. Dispositions are based on fuzzy logic which uses the whole interval between 0 (false) and 1 (true) to describe human reasoning. Fuzzy logic resembles human decision making and has an ability to draw a precise conclusion from approximate data.